\documentclass[preprint2]{aastex}
\begin{document}
\shortauthors{KLYPIN \& PRADA }
\shorttitle{Testing gravity with satellites}

\title{Testing gravity with motion of satellites around galaxies: Newtonian gravity against
 Modified Newtonian Dynamics}

\author
{Anatoly Klypin,${^1}$ and Francisco Prada$^2$}
\affil{$^1$Astronomy Department, New Mexico State University,
MSC 4500, P.O.Box 30001, Las Cruces, NM, 880003-8001, USA}

\affil{$^2$Instituto de Astrof\'\i sica  de 
Andaluc\'\i a (CSIC), Camino Bajo de Huetor, 50, E-18008 Granada, Spain}

\date{}

\newcommand\LCDM{$\char 3CDM$}
\def\simlt{\mathrel{\hbox{\rlap{\hbox{\lower4pt\hbox{$\sim$}}}\hbox{$<$}}}}
\def\simgt{\mathrel{\hbox{\rlap{\hbox{\lower4pt\hbox{$\sim$}}}\hbox{$>$}}}}
\def\ale{\mathrel{\hbox{\rlap{\hbox{\lower4pt\hbox{$\sim$}}}\hbox{$<$}}}}
\def\age{\mathrel{\hbox{\rlap{\hbox{\lower4pt\hbox{$\sim$}}}\hbox{$>$}}}}
\def\spose#1{\hbox to 0pt{#1\hss}}
\newcommand\lsim{\mathrel{\spose{\lower 3pt\hbox{$\mathchar"218$}}
\raise 2.0pt\hbox{$\mathchar"13C$}}}
\newcommand\gsim{\mathrel{\spose{\lower 3pt\hbox{$\mathchar"218$}}
\raise 2.0pt\hbox{$\mathchar"13E$}}}

\begin{abstract}
  The motion of satellite galaxies around normal galaxies at distances
  50-500~kpc provides a sensitive test for the theories.  We study the
  surface density and the velocities of satellites around isolated red
  galaxies in the Sloan Digital Sky Survey.  We find that the surface
  number-density of satellites declines with the projected distance as
  a power law with the slope $-1.5-2$. The rms velocities gradually
  decline: observations exclude constant velocities at a $\sim
  10\sigma$ level.  We show that observational data strongly favor the
  standard model: all three major statistics of satellites -- the
  number-density profile, the line-of-sight velocity dispersion, and
  the distribution function of the velocities -- agree remarkably well
  with the predictions of the standard cosmological model. Thus, that
  the success of the standard model extends to scales (50-500)~kpc,
  much lower than what was previously considered.  MOND fails on these
  scales for models which assume any single power-law number-density
  profile of satellites and any constant velocity anisotropy by
  predicting nearly constant rms velocities of satellites.  Satellite
  data can be fit by fine-tuned models, which require (1) specific
  non-power-law density profile, (2) very radial orbits at large
  distances (velocity anisotropy $\beta =0.6-0.7$ at $R=200-300$~kpc),
  and (3) 2-2.5 times more stellar mass than what is found in the
  galaxies. The external gravity force -- a necessary component for
  MOND -- makes the situation even worse.  We argue that a combination
  of satellite data and observational constraints on stellar masses
  make these models very problematic.
\end{abstract}

\section{Introduction}
\label{sec:intro}

One hundred years after Einstein, the theory of general relativity (GR) is
still our best theory of gravity. In the framework of GR, the standard
model of cosmology (\LCDM) provides a successful description of the
Universe. In this model, the same fluctuations which give rise to the
observed small variations in the temperature of the cosmic microwave
background (CMB) grow under the force of gravity, and eventually form
observed galaxies and other nonlinear structures such as filaments,
voids, groups and clusters of galaxies. According to the model, only
$\sim 4\%$ of the density in the Universe is provided by normal
baryonic matter \citep{WMAP06}. The \LCDM~ model requires two additional
components: a non-baryonic cold dark matter (CDM), which contributes
about 30\% of the average density of the Universe, and an even more
mysterious dark energy, which makes up the rest \citep{WMAP06}. The
model is remarkably successful on scales larger than a few
Megaparsecs. It predicted the amplitude and the spectrum of
angular fluctuations in the CMB and in the distribution of
galaxies \citep{Bardeen1987,Holtzman1989} that were later confirmed by
observations \citep{WMAP06,Boomerang2002,2dF,Tegmark2004}.
However, the \LCDM~ model faces challenges on
smaller scales. The most difficult ones are related with the
rotation in the inner parts of spiral galaxies. It seems
that the theory predicts too much dark matter inside
$\sim$~1kpc from the centers of galaxies \citep{Moore1994,Flores1994,deBlok2001}.
While there are some possible solutions of the
problem \citep{Rhee,Hayashi,Valenzuela}, the problems on small scales
are the
strongest challenge the standard model has encountered. When
compounded with the fact that there is no direct evidence of dark
matter or dark energy, the current problems of the standard
cosmological model have encouraged a small but growing community of
physicists to propose alternative theories of gravity to avoid the
need for dark matter.

This is the case for Modified Newtonian Dynamics (MOND), proposed by
 \citet{Milgrom} to explain the rotation of galaxies without dark
matter. According to MOND, the rotation curves in the outer regions
of galaxies do not decline because the force of gravity is
significantly stronger than for Newtonian gravity. 
At early times MOND's main appeal
was its simplicity: there is no need to make the assumption that the Universe is
filled with particles that nobody has seen. Additional motivation
came later from difficulties with explaining anomalies in the trajectories
of the Pioneer 10 and 11 space missions \citep{Pioneer}.

Yet, for a long time MOND was not more than a conjecture. Only recently,
Bekenstein proposed a relativistic version named tensor
vector scalar theory (TeVeS) \citep{Bekenstein2004}. This alternative theory of gravity
provides a framework to make predictions of numerous important
observational phenomena, which \LCDM~ has already done: the temperature
fluctuations seen in the CMB, gravitational
lensing, and the large scale structure of the universe. With maturity
came problems. Rotation curves of some galaxies -- the
initial strong argument for MOND -- cannot be explained by MOND.
In about 1/4 of galaxies considered by proponents of MOND the predicted velocities
are well above
the observations in the very central regions \citep{Sanders2002}.
RMS velocities of stars in some dwarf spheroidal galaxies \citep{Lokas2006}
also present problems.

So far, the most severe challenges for MOND are coming from clusters
of galaxies. Dynamics of galaxies in clusters cannot be explained by
MOND and requires introduction of dark matter, possibly in the form of
a massive ($\sim 2$eV) neutrino \citep{Sanders2002}. We do not know
whether this modification can explain some properties of clusters of
galaxies such as the ``Bullet Cluster'', where the baryonic mass
(galaxies and gas) is clearly separated from the gravitational mass,
as indicated by gravitational lensing \citep{Clowe2006,Angus2006}. In
any case, for MOND to survive too it must invoke dark matter in the
form of massive neutrinos and dark energy in the form of an arbitrary
constant added to a combination of two scalar fields used in TeVeS
MOND \citep{Bekenstein2004}.

There is no doubt that alternative theories of gravity represent a
challenge to the standard model of cosmology and GR. Any theory or
model must respond to these challenges. Here we present a number of
observations to test gravity and dark matter in the peripheral parts
of galaxies at distances 50-500~kpc from the centers of
galaxies. These scales can be tested by studying the motion of
satellites of galaxies.  This is a relatively old field in
extragalactic astronomy and historically it was one of main arguments
for the presence of dark matter \citep{ZW1994,Prada2003}.

The paper is organized as follows. In Section 2, we present the
observational results drawn from the SDSS and the predictions from the
standard model of cosmology. Predictions from MOND are computed and
discussed in Section 3. Finally, conclusions are given in Section 4.

\begin{figure*}[tb!]
\epsscale{2.0}
\plotone{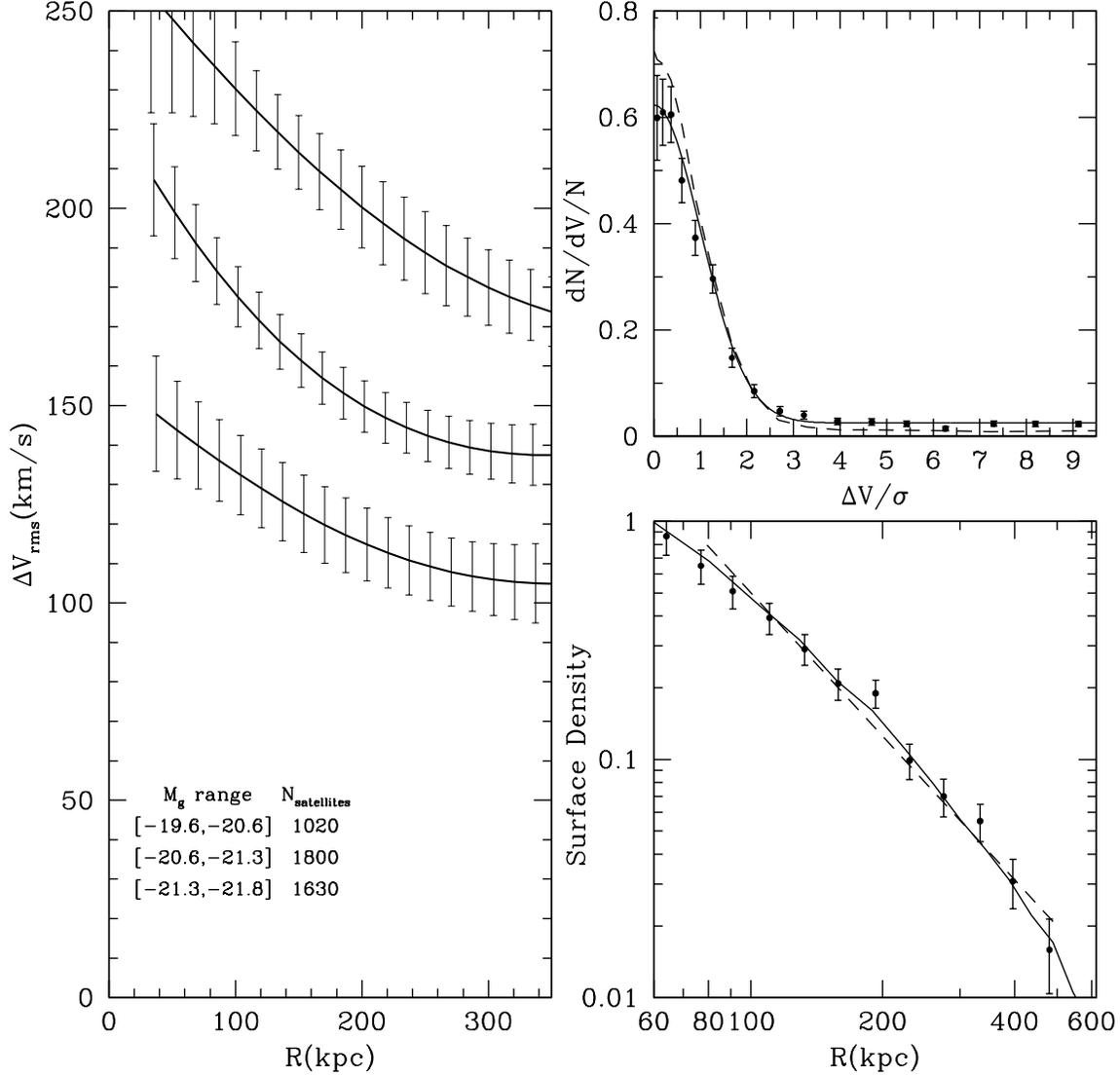}
\caption{{\it Left panel:} RMS of the line-of-sight velocities of
  satellites orbiting red isolated galaxies with absolute magnitudes
  indicated in the plot. The luminosity decreases from top to the
  bottom curve. Satellites move faster around more luminous
  galaxies. In all samples of galaxies the rms velocity is declining
  with distance. {\it Right top panel:} Distribution of observed
  line-of-sight velocities (dots with error bars) has simple structure
  of the Gaussian distribution with a small constant background (full
  curve). The background is due to objects, which in projection lie
  close to the central galaxy, but that are far from it in 3D
  space. The dashed curve is the distribution of velocities expected
  in the standard cosmological model drawn from \LCDM~
  simulations. {\it Right bottom panel:} Surface number-density of
  satellites (dots with error-bars; arbitrary units) orbiting galaxies
  with luminosity $-20.0 < M_g < -21.5$. The dashed line is the power
  law with a slope $-2$. The full curve is the prediction of the
  \LCDM~ model.}
\end{figure*}

\section{Observational results}
\label{sec:results}

We use the Sloan Digital Sky Survey (SDSS; www.sdss.org) -- the
largest photometric and spectroscopic astronomical survey ever
undertaken of the local Universe -- to study the motion of
satellites. As of Data Release Four (DR4) \citep{dr4}, imaging data
are available over 6670~deg$^2$ in five photometric bands. In addition
to the CCD imaging, the SDSS 2.5m telescope on Apache Point, New
Mexico, measured spectra of galaxies, providing distance
determinations.  Approximately half million of galaxies brighter than
$r=17.77$ over 4700~deg$^2$ have been targeted for spectroscopic
observations as part of SDSS and are included in DR4. Redshift
accuracy is better than 30~km/s and the overall completeness is
$\sim$90\%.  For our study we compute rest frame absolute magnitudes
in the g-band from the extinction-corrected apparent magnitudes
assuming a $\Lambda CDM$ cosmology with a Hubble constant $h=0.7$
($H_0$ = $100 h {\rm km s}^{-1} {\rm Mpc}^{-1}$).  Galaxies are split
into red (early-types) and blue (late-types) populations based on the
bimodality observed in the $u-r$ color distribution
\citep[e.g.]{Baldry2004}.  The local minima between the peaks of the
color distribution occur near $u-r=2.3$. All magnitudes and colors are
k-corrected to $z=0$. Because calculations of MOND gravity for
non-spherical objects are complicated, we restrict our analysis only
to red galaxies, the vast majority of which are either elliptical
galaxies or are dominated by bulges.  Our galaxy sample was selected
from the full redshift sample by taking all galaxies with recession
velocity 3000~km/s~$< cz < 25000$~km/s. The total number of selected
galaxies is about 215,000. The SDSS heliocentric velocities were
converted to the Local Group standard of rest before computing
distances.  We select our host galaxies as galaxies with absolute
g-band magnitude brighter than $M_g= -19.0$ and isolated: a galaxy
must be at least 4 times brighter than any other galaxy within a
projected distance $R < 1$~Mpc and a line-of-sight velocity difference
$\Delta V< 1500$~km/s.  We define satellites as all galaxies being at
least 4 times fainter than their hosts and found within a projected
distance $R < 1000$~kpc and velocity difference $\Delta V < 1500$~km/s
with respect to their hosts.  Typically we find about 1.5 satellites
per host.  In total we have 9500 satellites with a mean luminosity of
about $M_g= -18.0$. We bin the host galaxies
by luminosity and collect information about the distribution of
relative velocities $\Delta V$ and the number of satellites as the
function of projected distance $R$. 

Figure~1 presents the observational results for primaries, which are
at least 6 times brighter than any satellite. We also used primaries,
which are 4 times brighter, and primaries, which are 10 times brighter
than their satellites. We do not find any trend with the
primary-satellite magnitude gap: results are nearly the same (see
Section 3 for details). The only difference is the statistics, and,
thus, the error-bars. The distribution of line-of-sight velocities
(top right panel) clearly shows a two-component structure: a
homogeneous background of interloper galaxies (dwarfs which happen to
lie along the line-of-sight, with large physical distances from the
host galaxies but small projected and velocity differences, but which
are not associated with the host) and a nearly Gaussian component. The
surface density of the satellites also shows the same structure: at
large separations the number density goes to a constant due to
interlopers.  We subtract the background and plot the surface density
of satellites in the bottom right panel.  In order to study the
velocity dispersion of satellites $\Delta V_{\rm rms}^2\equiv \langle
\Delta V^2 \rangle$ as a function of projected distance $R$ we use a
maximum likelihood method to approximate the number of satellites
$N(R,\Delta V)$ per unit projected area and per unit velocity
difference $\Delta V$ using an 9-parameter function in the form of a
constant plus a Gaussian distribution with variable velocity
dispersion and normalization:
\begin{equation}
 {N(R,\Delta V)} =
 n(R)\exp\left[-{V^2}/{2\sigma^2(R)}\right]/\sqrt{ 2\pi}\sigma+ n_0,
\label{eq:Delta}
\end{equation}
where the surface number-density of satellites $n(R)$ and the rms
line-of-sight velocity $\sigma=\Delta V_{\rm rms}(R)$ are 3th order
shifted Chebyshev polynomials of the first kind.  The parameter $n_0$
is a constant representing the background of interlopers. The left
panel in Figure~1 shows the resulting rms velocity of satellites for
three magnitude bins. In order to estimate statistical uncertainties,
we run Monte Carlo simulations using the same number of hosts and
satellites as in the corresponding magnitude bin in Figure~1. Note
that the data points are correlated. We use Monte Carlo simulations to
test statistical significance that the observed velocities are
declining with distance: data for each magnitude bin reject a constant
$\Delta V_{\rm rms}(R)$ at about $3\sigma$ confidence level. We also
studied blue galaxies and find declining velocities for them. These
results are in agreement with previous
estimates \citep{Prada2003,Conroy2007}, but we now can exclude constant
rms velocities much more reliably.

\begin{figure*}[tb!]
\epsscale{2.0}
\plotone{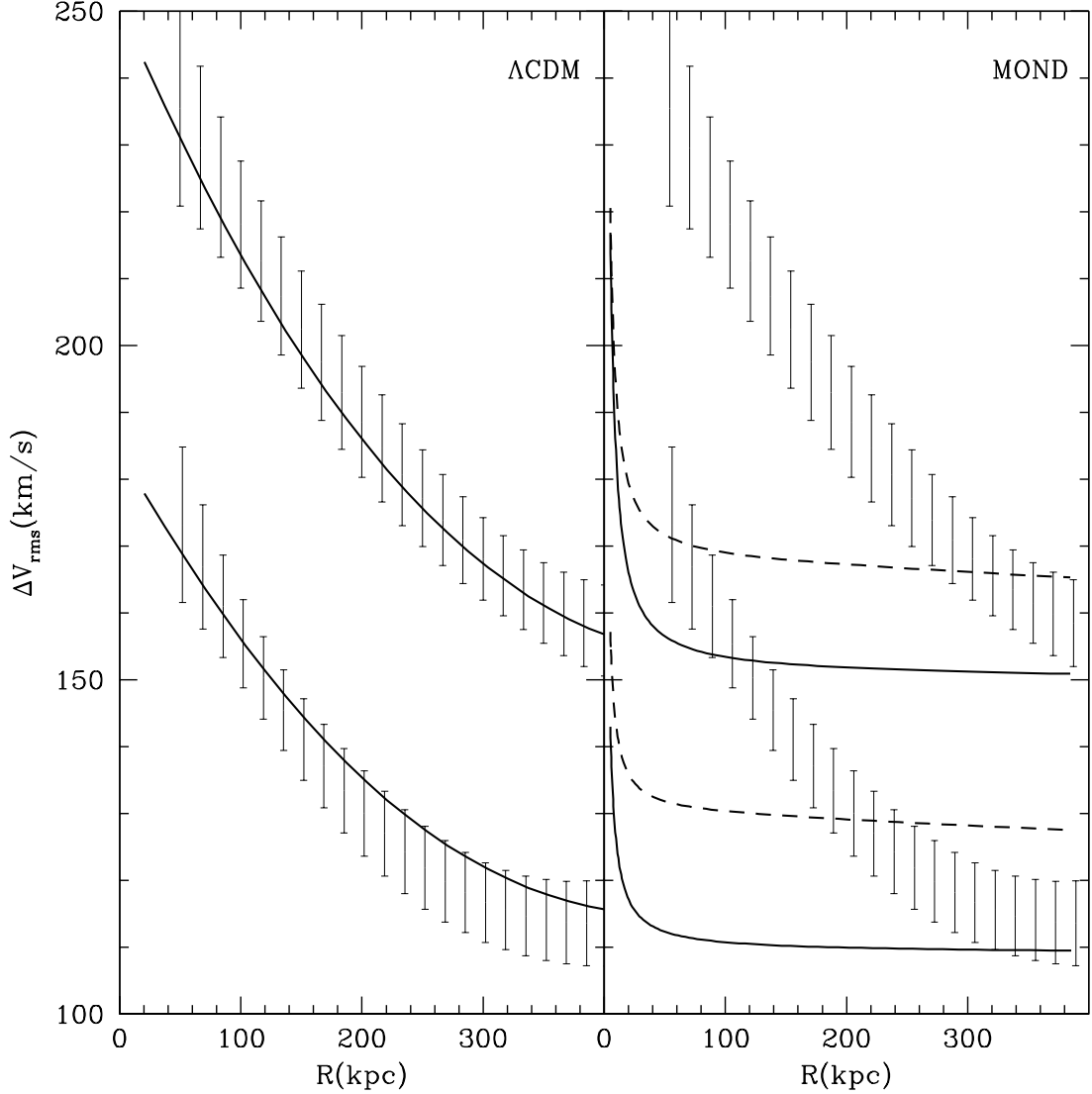}
\caption{The line-of-sight velocities of galaxies in two luminosity
  ranges: $-20.5 < M_g < -21.1$ (2300 satellites) and $-21.1 < M_g <
  -21.6$ (2700 satellites). The vertical lines indicate $68\%$
  confidence levels.  The left panel shows predictions (full curves)
  of the standard cosmological model for galaxies hosted by dark
  matter halos with maximum circular velocities of $\sim 340$~km/s
  (top curve) and $\sim 270$~km/s (bottom curve). The right panel
  shows MOND predictions. The two bottom curves are for stellar masses
  $8\times 10^{10}M_{\odot}$ estimated for galaxies with
  $M_g=[-20.5,-21.1]$. The full curve is for models with isotropic
  velocities $\beta=0$ and with the observed slope $\alpha=-3$ of the
  3D number density of the satellites. The curve was so much below the
  observed data points that for other models we decided to use the
  slope $\alpha=-2.5$, which is marginally compatible with the
  observations. The dashed curve is for orbits, which are
  preferentially radial ($\beta={\rm const}=0.5$). The two top curves
  are for the stellar mass $2\times 10^{11}M_{\odot}$  and for $\beta=0$ (full curve) and
  $\beta=0.6$ (dashed curve). The \LCDM~ makes quite reasonable
  predictions, while MOND has problems.}
\label{fig:two}
\end{figure*}

\section{Predictions of the \LCDM model}
\label{sec:resultsLCDM}

In order to compare observational results with the \LCDM~ predictions,
we use high-resolution cosmological N-body simulations
\citep{Prada2006}, which we then treat as if they are the
observational data.  The simulations were done for the standard \LCDM
model with parameters $sigma_8=0.9$, $\Omega_0=0.3$, $h=0.7$,
$n=1.$. We use two simulations. Both simulation had $512^3$
particles. One simulation had the computational box 120~$h^{-1}$Mpc
box, mass resolution $m_1=1.07\times 10^9h^{-1}M_\odot$ and force
resolution 1.8~$h^{-1}$kpc. The second simulation has the
computational box 80~$h^{-1}$Mpc box, mass resolution $m_1=3\times
10^8h^{-1}M_\odot$ and force resolution 1.2~$h^{-1}$kpc. We do not
find any differences between the simulations and decided to present
results of the larger simulation, which provides better statistics of
halos. We select isolated halos as halos, which within projected
distance of 715~kpc and within relative velocity difference 1000~km/s
do not have halos or subhalos with maximum circular larger than 1/2 of
the halo's circular velocity. This corresponds to the mass ratio of a
factor ten for distrinct halos. Circular velocities are better
characteristics of subhalos, which have poorly defined
masses. Although we resolve subhalos, we dicided to use dark matter
particles as proxies for satellites. At distances larger than a
fraction (1/4- 1/3) of a virial radius from center of a halo
satellites trace the motion and the spacial distribution of the dark
matter \citep[e.g.][]{Sales2007}. 

Results presented in Figures 1 and 2 show that all three
characteristics of observed satellites are reproduced by the
model. Note that we actually do not make fits.  Results from
simulations have only one parameter: the maximum circular velocity of
the dark matter halo. We just plot what we obtain from the
simulations. 

Once we fix the maximum circular velocity, the results
are fairly insensitive to parameters of the cosmological model. Two
factors may affect the shape of the theoretical V-R diagram: the halo
concentration and the velocity anysotropy. The halo concentrations
$c\equiv R_{\rm vir}/R_s$ are in the range $c=8-10$ for halos ($M_{\rm
  vir}=(2-10)\times 10^{13}M_\odot$) and normalization $\sigma_8$
cosidered in this paper \citep{Bullock2001,Neto2007}. (Here $R_{\rm
  vir}$ is the virial radius and $R_s$ is the characteristic radius of
the NFW profile). For virial radii in the range $300-500$~kpc, the
exact value of the halo concentration affects only the central region
$R<50-70$~kpc. The average halo concentration depends weakly on the
amplitude of the perturbations $\sigma_8$. Existing observational data
leave only very narrow range for variation of $\sigma_8=0.8-0.9$,
which also limits the range of concentrations.  The velocity
anisotropy is less certain, but it is not large. N-body simulations
indicate that $\beta$ is slightly positive and has a tendency to
increase with the distance for radii smaller than the virial radius
\citep[e.g.][]{Cuesta2007} with typical values $\beta\approx 0.2-0.3$
for halo masses considered in this paper. At larger distances $\beta$
declines and goes to zero \citep{Cuesta2007}.

\begin{figure}[tb!]
\epsscale{1.25}
\plotone{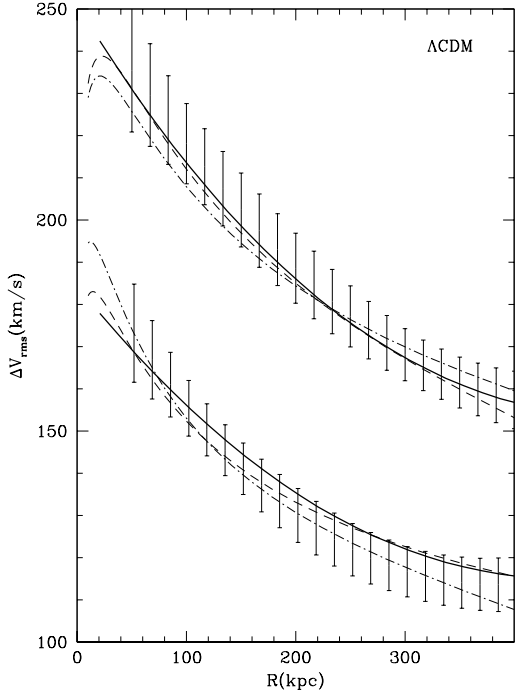}
\caption{
  Comparison of results of N-body simulations (full curves) with
  solutions of the Jeans equation (dashed and dot-dashed curves. In
  the top part of the plot the vertical error bars show observational
  constraints for galaxies in the luminosity range $-21.1 < M_g <
  -21.6$. The corresponding full curve is for halos in simulations
  with maximum circular velocity $\sim 340$~km/s. The two other curves
  are solutions of stationary spherically symmetric Jeans equation
  with the same maximum circular velocity. The dashed curve is for
  slightly more radial velocities (velocity anisotropy $\beta=0.34$ at
  $R=300$~kpc. The dot-dashed curve of for a model with $\beta=0.17$
  at the same distance. The bottom curves are for halos with maximum
  circular velocity $\sim 270$~km/s. The vertical error bars are for
  galaxies in the luminosity range $-20.5 < M_g < -21.1$. The other
  curves are labeled in the same way as  the top curves.
\label{fig:Jeans}
}
\end{figure}

In order to demonstrate that our results of N-body simulations are
robust and reliable, we use solutions of the Jeans equation, which we
obtain for parameters compatible with numerous previous simulations.
Once a solution for the radial velocity dispersion is obtained, we
integrate it along a line of sight with appropriate corrections for
the velocity anisotropy. For the density use the NFW profiles with
concentration $c=9-10$. For the sake of completeness, we add the
cosmological background density, which has only a small effect at
large distances. For $\beta$ we use approximation given by
eq.~(\ref{eq:betb}) with parameters chosen in such a way that
$\beta(0)=0$ and at $R=300$~kpc $\beta =0.2-0.4$. We use the same
range of circular velocities as in Figure~\ref{fig:two}. For more
massive halos we use $M_{\rm vir}=8.5\times 10^{12}h^{-1}M_\odot$ and
$c=9$. For less massive halos we adopted $M_{\rm vir}=3.8\times
10^{12}h^{-1}M_\odot$ and $c=10$. Results presented in
Figure~\ref{fig:Jeans} clearly indicate that very simple stationary
models can accurately reproduce results of simulations.

\section{Predictions from MOND}

\subsection{Analytics}
The situation is different for MOND because there are no predictions
on the same level of sophistication as for \LCDM. In principle, those
predictions can be made, but at this moment they have not been yet
made.  Thus, we have only one option:  solve the Jeans
equation for spherical systems. When doing so, we have a freedom of
choosing two functions: the number-density profile of satellites and
the velocity anisotropy $\beta(r)$.  We also have two free parameters:
stellar mass $M_*$ and the magnitude of external force (see below for
details).  The predictions are constrained by two functions $n(R)$ and
$\Delta V(R)$. The velocity distribution function also gives
constraints, but those are relatively weak. Only the models with a
large $|\beta|> 0.85$ can be excluded. The stellar luminosity and
colors constrain the stellar mass. Roughly speaking we have two
arbitrary functions to fit two observed functions. It should not be
difficult given that the model is viable.

In our case the Jeans equation for the radial velocity dispersion
$\sigma^2(r)$ can be written in the form:

\begin{equation}
\frac{d\sigma^2}{dr} + \sigma^2\frac{(2\beta+\alpha)}{r} =-g(r), 
     \alpha\equiv \ln\rho/d\ln r,
\label{eq:Jeans}
\end{equation}
where $\beta(r)=1-\sigma^2_{\perp}/2\sigma^2$ is the velocity
anisotropy and $g(r)$ is the gravitational acceleration. The formal
solution of the Jeans equation can be written in the form:

\begin{equation}
    \sigma^2 = \frac{1}{\chi (r)\rho(r)}\int_r^{\infty}\chi (r)\rho(r)g(r)dr,
\label{eq:JeansSol}
\end{equation}

\begin{equation}
 \chi(r) =  \exp\left[2\int^r\frac{\beta dr}{r}\right].
\label{eq:JeansChi}
\end{equation}

\begin{figure*}[tb!]
\epsscale{2.2}
\plottwo{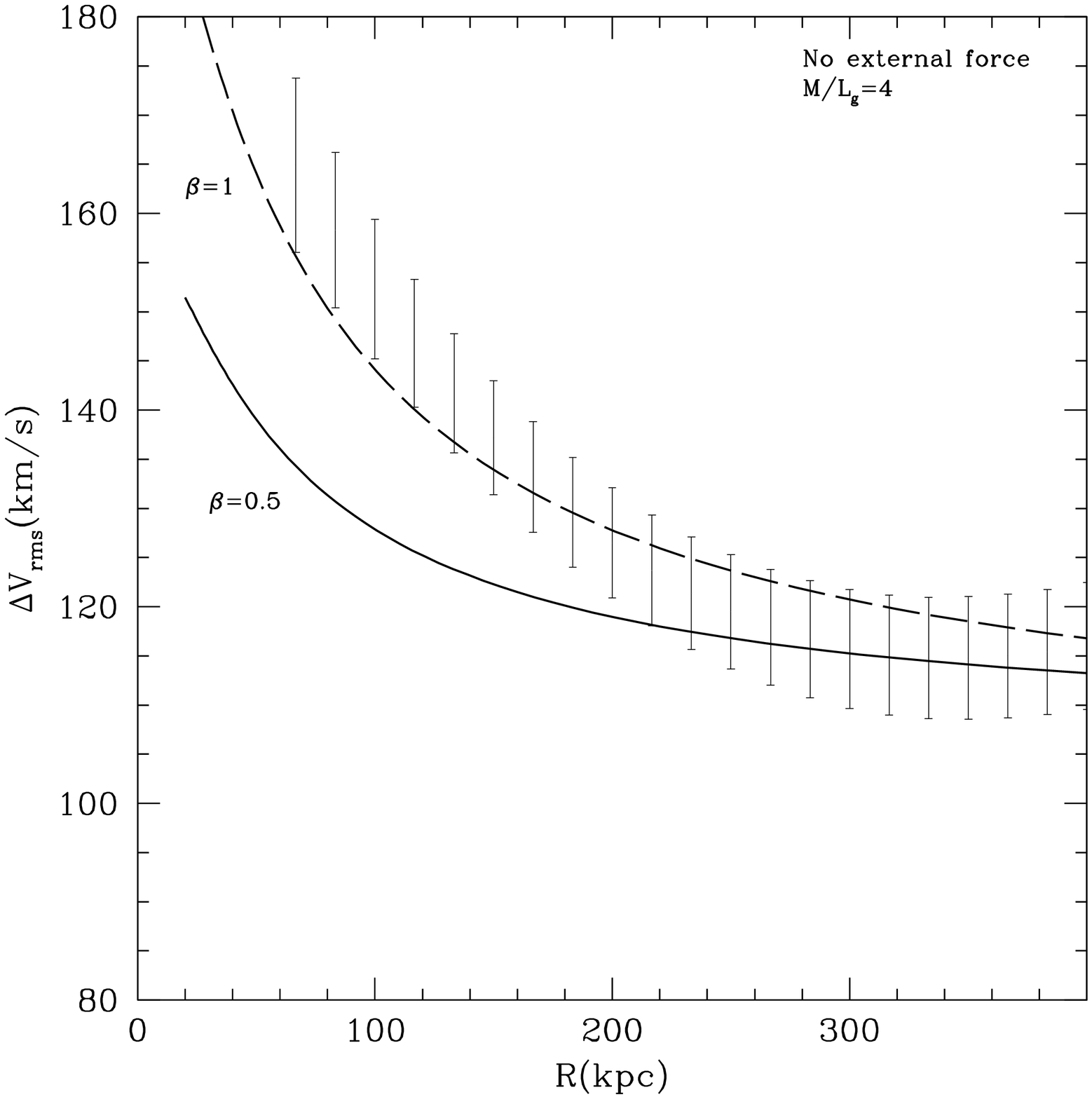}{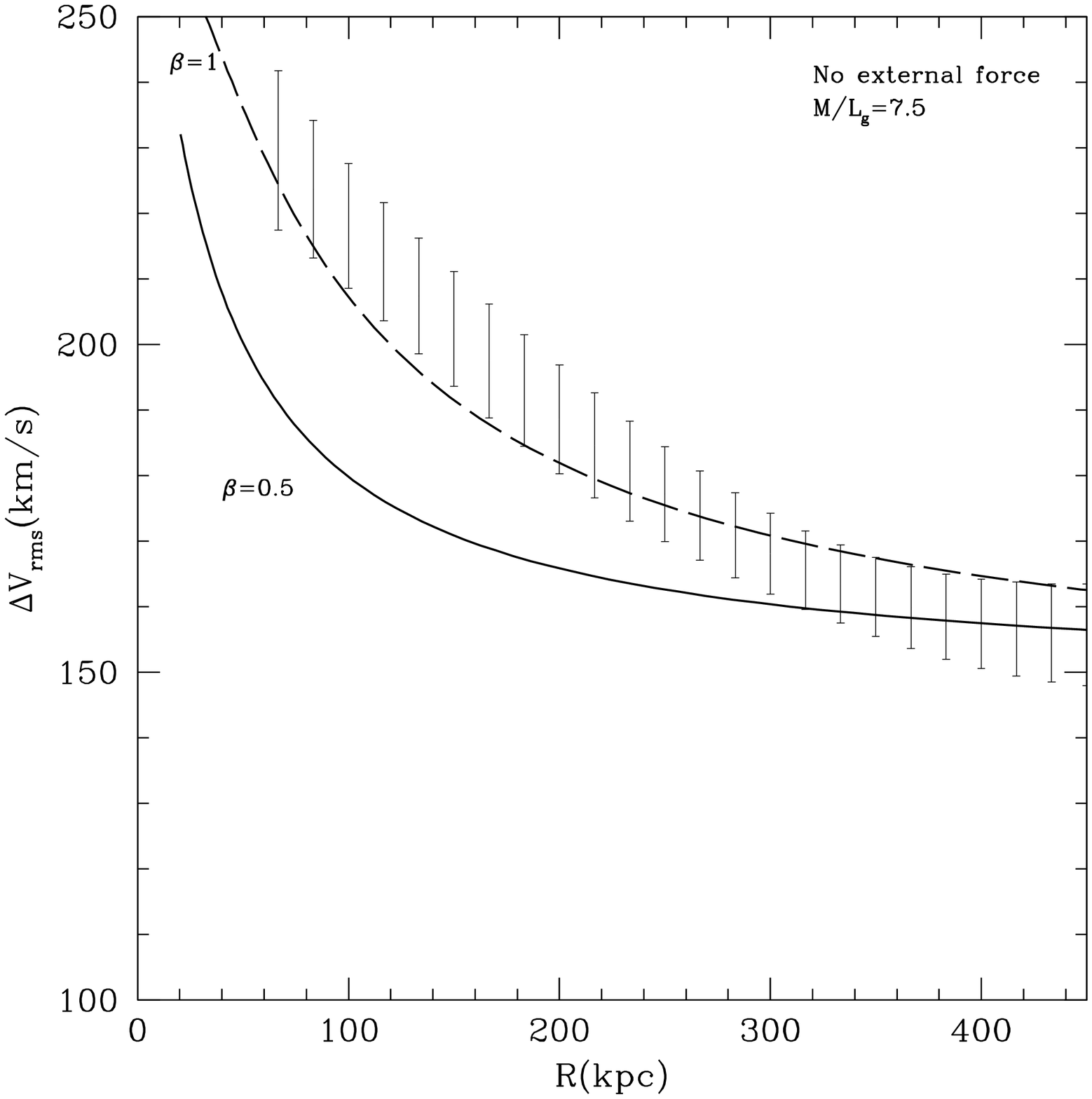}
\caption{
The line-of-sight velocities of galaxies with $-20.1 < M_g < -20.9$
(left panel) and $-21.1 < M_g < -21.6$ (right panel) are compared with
MOND models with carefully tuned parameters.  We use $M_*=8\times
10^{10}M_{\odot}$ and $M_*=3\times 10^{11}M_{\odot}$ for the two
bins.  No external gravity force was assumed. Models of MOND
can fit the data only when very radial orbits with $\beta_0=1.0$
(eq.(7)) are used (dashed curves). Models with more reasonable
$\beta_0=0.5$ fail (full curves). Data for more luminous galaxies (right panel) 
require twice as much of stellar mass as actually observed in the galaxies.
\label{fig:Tuned}
}
\end{figure*}

It is convenient to chose forms of $\alpha$ and $\beta$ such that the
integral in eq.(\ref{eq:JeansChi}) is taken in elementary functions. We
used the following approximations:

\begin{equation}
   \rho = r^{-\alpha}(1+\frac{r}{r_a})^{-\alpha_1},
\label{eq:alpha}
\end{equation}

\begin{equation}
   \beta = \beta_0,  \quad \chi = r^{2\beta_0},
\label{eq:beta}
\end{equation}
\begin{equation}
       \beta = \beta_0 +\frac{\beta_1}{1+\frac{r}{r_b}},  \quad
       \chi = r^{2\beta_0}\left[\frac{r}{r_b}\frac{1}{(1+\frac{r}{r_b})}\right]^{2\beta_1}
\label{eq:betb}
\end{equation}

\begin{equation}
     \beta = \beta_0 +\frac{\beta_1}{1+(\frac{r}{r_b})^2},
        \chi = r^{2\beta_0}\left[\frac{r}{r_b}\right]^{2\beta_1}
                    \left[\frac{1}{1+(\frac{r}{r_b})^2}\right]^{\beta_1}    
\label{eq:betc}
\end{equation}

In the case of Newtonian gravity the acceleration
$g=g_N(r)=GM(r)/r^2$, where the mass $M(r)$ includes both normal
baryonic mass and dark matter. For MOND the acceleration of spherical
systems $g_{\rm MOND}$ is given by the solution of the non-linear
equation
\begin{equation}
  g_{\rm MOND}\mu(|{\bf g}_{\rm MOND}+{\bf g}_{\rm ext}|/a_0)=GM_*(r)/r^2,
 \label{eq:gmond} 
\end{equation}
where $M_*$ is the mass of only baryons and $a_0=1.2\times 10^{-8}{\rm
  cm}\: {\rm sec}^{-2}$. The term $\bf{g}_{\rm ext}$ is the external
constant gravitational acceleration. It formally breaks the spherical
symmetry. Thus, the eq.~(\ref{eq:gmond}) is only an approximation
valid for small external force. Here we use the same approximation as
eq.~(10) in \citet{Sanders2002}, which we average over angles between
a constant external acceleration $\bf{g}_{\rm ext}$ and internal
radial acceleration $g_{\rm MOND}$. 

The function $\mu(x)$ can have different shapes. We tried the originally
proposed form \citep{Milgrom} $\mu(x)=x/\sqrt{1+x^2}$, but accepted and used
 for all our analysis the function $\mu(x)=x/(1+x)$, which gives slightly better results.
The function $\alpha(r)$ is limited by the observations presented in
Figure~1.  The velocity anisotropy $\beta(r)$ is a free function, but
there are constraints.  Asymptotically $\beta$ goes to zero at large
distances where gravitational effect of the central galaxy
diminishes. Ideally the velocity anisotropy at small distances also
should be declining. There are different arguments why this should be
the case. (1) The tangential velocities of few satellites of the Milky
Way, for which the proper motions are measured, strongly reject radial
orbits \citep{Kallivayalil2006, Piatek2007}. There is no reason to
believe that our Galaxy should be special in this respect. (2)
Experience with gravitational dynamical systems indicate that in
dynamically relaxed systems $\beta \approx 0$. Numerous simulations of
cosmological models illustrate this: $\beta$ is small in the central
region and increases to $0.2-0.3$ at the viral radius \citep{Cuesta2007}. Note that we
should distinguish the velocity anisotropy and the orbital
eccentricity. For centrally concentrated objects, which we are dealing
with here, already isotropic velocities imply typical peri-/apocenter
ratios of 1:4-1:5. If eccentricity is larger, a significant fraction
of satellites comes too close to the central $\sim 10$~kpc region
where the satellites are destroyed by tidal forces.

 The term $\bf g_{\rm ext}$ in eq.(\ref{eq:gmond}) describes the
 effect of external gravitational field. It is specific for MOND. In
 Newtonian dynamics a homogeneous external gravity does not affect
 relative motions inside the object.  Because the MOND gravity is
 nonlinear, the internal force is affected by the external field (note
 that this is not the tidal force). This external effect is quite
 complicated.  The magnitude of the
 external force is substantial for the motion of the
 satellites. \citet{Fam2007,Wu2007} point out that in MOND numerous
 sources and effects generate about equal magnitude of $g_{\rm ext}$.
 For example, 600~km/s motion of the Local Group relative to the CMB
 implies acceleration of 600~km/s$/10$~Gyrs$=0.015a_0$. Here we assume
 that the acceleration is constant over the whole age of the Universe.
 If $g_{\rm ext}$ increases with time, as it may be expected, then the
 acceleration is even larger. Infall with $250$~km/s in the direction
 of Virgo gives about half of the value. M31 produces about the same
 magnitude of the acceleration.  While the acceleration in MOND does
 not add linearly, it is reasonable to assume that $g_{rm
   ext}>0.015a_0$. We will explore the effect the external field
 later.

 \subsection{Comparison of MOND models with observations}
 In order to make MOND predictions, we must estimate the stellar mass
 for galaxies in our analysis.  For two subsamples presented in
 Figure~\ref{fig:Tuned} we use the magnitude bins $M_g =-20.5-21.1$
 (2400 satellites) and $M_g =-21.1-21.6$ (2700 satellites). The
 average luminosities of galaxies in the bins are $L_g=2.0\times
 10^{10}L_{\odot}$ and $L_g=4.0\times 10^{10}L_{\odot}$.  When
 estimating the luminosities, we assume $M_{g,\odot} =5.07$
 \citep{Blanton01}.  Using measured $u-g$ colors we estimate the
 stellar masses of galaxies \citep{Bell01}: $M_*= 7.2\times
 10^{10}M_{\odot}$ and $M_*= 1.5\times 10^{11}M_{\odot}$ for the two
 luminosity bins.  This implies the mass-to-light ratios are nearly
 the same for the bins: $M/L_g=3.7$.  These $M/L$ estimates are close
 to predictions of stellar population models. In our case, the
 galaxies are red ($u-g >2.3$), old mostly ellipticals, for which we
 expect nearly solar metallicity. Indeed, \citet{Maraston05} gives
 $M/L_B=5$ for the Kroupa IMF for stellar population 10~Gyrs
 old. Adjusting for the 0.4 mag difference between B and g bands, we
 get $M/L_g=3.5$ (the Salpeter IMF gives $M/L_g=5.5$).  Larger
 metallicity is highly unlikely (the galaxies are not really massive
 ellipticals) and it does not make much difference: 10 percent
 increase if we take twice the solar metallicity.

\begin{figure}[tb!]
\epsscale{1.0}
\plotone{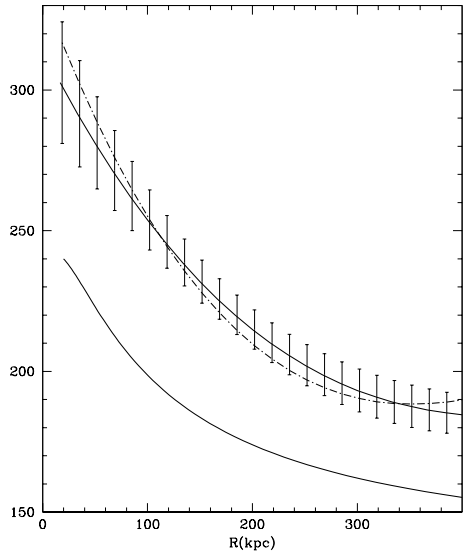}
\caption{
 RMS velocities of satellites for galaxies selected by stellar mass
with the average $M_*=2.2\times 10^{11}M_\odot$. The top full curve
with the error bars is for a sample of 2600 satellites, which are more
than 6 times dimmer than the primary galaxy. The dashed curve is for
more stringent isolation criterion of satellites 10 times less bright
than the primary (995 satellites). The differences are not
statistically significant and are less than 10~km/s for
$R>50$~kpc. The bottom curve is for MOND with the same stellar mass as in the observations
and with tuned parameters to produce the best fit. The MOND prediction
 falls much below the observational data.
\label{fig:Stmass}
}
\end{figure}

\begin{figure}[tb!]
\epsscale{1.0}
\plotone{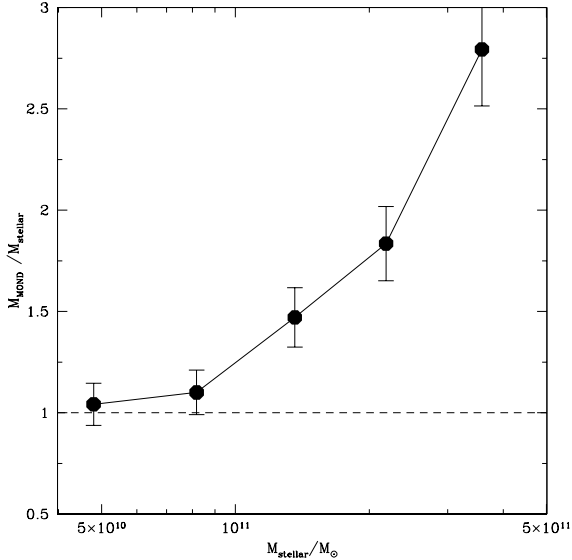}
\caption{The ratio of mass required by MOND to the stellar mass for galaxies selected  by stellar mass.
For small mass galaxies MOND requires masses, which are in reasonable agreement with the observed 
stellar masses. Yet, for more massive galaxies  MOND needs more mass than actually is observed.
\label{fig:MONDmass}
}
\end{figure}

 Using different functional forms for $\beta(r)$ , we
solve the equations numerically and then integrate the solution along
the line-of-sight to get a prediction for $\Delta V_{\rm
rms}$. Figure~2 presents results for the two different magnitude ranges and
for different parameters $\beta=$const and $\alpha=$const.
MOND definitely has problems fitting the data because for any constant $\beta$ it predicts a
sharply declining rm velocity at small distances followed by nearly
flat velocities at large distances. The data show just the opposite
behavior. It is easy to understand why predictions of MOND has this
shape.  At large distances the Newtonian acceleration is very small
and the acceleration is strongly dominated by the MOND correction:
$g_{\rm MOND}\propto 1/r$.  In this case, the solution of eq.(1) gives
$\sigma=$const for {\it any} constant value $\beta$ and $\alpha$. 

We can try to salvage MOND, but so far there is no simple way of doing
it.  \citet{Angus2007} find that the way to improve the situation is
to have a model with variable slope $\alpha$ and with extremely radial
orbits. We also tried different functional forms and numerous
combinations of parameters.  While the parameters, which fit the
surface density and the rms velocities vary, they all give the same
answer: density slope in the central 40~kpc should be $\alpha >-2.0$
and the velocity anisotropy at large distances should be $\beta>0.7$.
Figure~\ref{fig:Tuned} shows (dashed curves) the results for the same set of
parameters as in \citet{Angus2007}: $\alpha=0$, $\alpha_1=3.1$, $r_a=40$~kpc,
$r_b=(20-40)$~kpc, $\beta_0 =1$ and $\beta_1=-2.57$. Here we use the
approximation given in eq.(\ref{eq:betb}).  The solution is very
contrived: relatively small deviations from the best behavior (small
slope in the center and radial orbits in the outer radii) result in
failed fits.  The full curves in the Figure show what happens when the
velocity anisotropy gets less radial: $\beta_0 =0.5$.

Even this fine-tuned solution fails unless the mass-to-light ratio for
the larger magnitude bin is arificially increased by a factor of two:
$M/L_g=7.5$, which gives $M_*=3\times 10^{11}M_{\odot}$.
\citet{Angus2007} also found the same trend, but they made two
mistakes, which did not allow them to clearly see the problem. First,
the solar mass-to-light ratio was used for the B band instead of the g
band. Second, the width of the magnitude bin is presented as an error
in M/L giving impression of very large uncertainties in M/L. This is
not correct: the statistical uncertainty of the average luminosity of
galaxies in each bin is very low and can be neglected. It should be
noted that there is nothing special about the galaxies, which are used
here. The average luminosities differ by a factor 1.7. So, it is not a
large difference. Colors of the galaxies are practically the same,
which then gives the same M/L if we use stellar population models. The
rms velocities are also perfectly consistent with simple scaling. For
example, the ratio of rms velocities at the same projected distance of
70~kpc is 1.3 implying simple scaling $L\propto \sigma^2R$. Roughly
speaking, we double the luminosity and that doubles the stellar
mass. This does not work for MOND: it needs twice more stellar mass.
This is definitely a problem because there is no justification why
galaxies with the same colors, with the same old population and
practically the same luminosity should have dramatically different
IMF. The differences are very large: most of the stellar mass in the
more luminous bin should be locked up in dwarfs with $\sim
0.2M_{\odot}$, while there is relatively little of those in the lower
bin, which is consistent with the Kroupa IMF.
 
In order to make the argument even more clear, we make analysis of
velocities in a different way. This time we split the sample by
stellar mass, but we still keep only red primaries with $u-r >2.3$. We
make the analysis twice: for satellites 4 times and for satellites 10
times less bright than the primary galaxy. There is no systematic
difference between the two isolation conditions: within $1\sigma$ the
results are the same.  Figure~\ref{fig:Stmass} illustrates this
point. We select primary galaxies to have the stellar mass in the
range $M_*=(1.6-3.2)10^{11}M\odot$. The average stellar mass is
$\langle M_*\rangle =2.2\times 10^{11}M\odot$, and the average
luminosity is $\langle L_g\rangle =5.4\times 10^{10}l_\odot$.  There
is a hint that more isolated primaries have slightly {\it larger}
velocities of satellites. Still, the differences are not statistically
significant: for radii larger than 50~kpc the differences are smaller
than 10~km/s. Then we take the observed stellar mass and use it for
MOND models and apply the best tuned parameters.  We find that
parameters suggested by \citet{Angus2007} ($r_b=40$~kpc, $\beta_0=1$,
$\beta_1=-2.57$) improve the fits as compared with a constant $\beta$
models. Still, they are not acceptable. For example, at $R=150$~kpc
the MOND model is a $7\sigma$ deviation. We found a better solution
($r_b=30$~kpc and other parameters the same), which places MOND
``only'' at $6.2\sigma$.

In order to envestigate what stellar mass is needed for MOND, we split
the sample of red primaries into 5 mass bins ranging from $M_*\sim
5\times 10^{10}M_\odot$ to $M_*\sim 5\times 10^{11}M_\odot$.  We chose
the less stringent isolation condition because it gives typically
2-2.5 time more satellites resulting in smaller statistical errors.  Each bin has a
large number of satellites: $1200-2900$. We then run a grid of MOND
models with different masses and select models, which make best fits.
Just as the MOND models in Figure~\ref{fig:Tuned}, there are models,
which marginally fit the data. For each model we get stellar mass
required by MOND and compare it with the stellar mass estimated by the
stellar population models. Results are presented in
Figure~\ref{fig:MONDmass}.  For smaller masses MOND gives masses,
which are compatible with actual stellar mass observed in the
galaxies.  The plot clearly shows the problem with massive galaxies:
MOND requires increasingly more mass than observed in the galaxies
ending up in large ($\sim 2.5$) disagreement with the observations.

\begin{figure*}[tb!]
\epsscale{2.4}
\plottwo{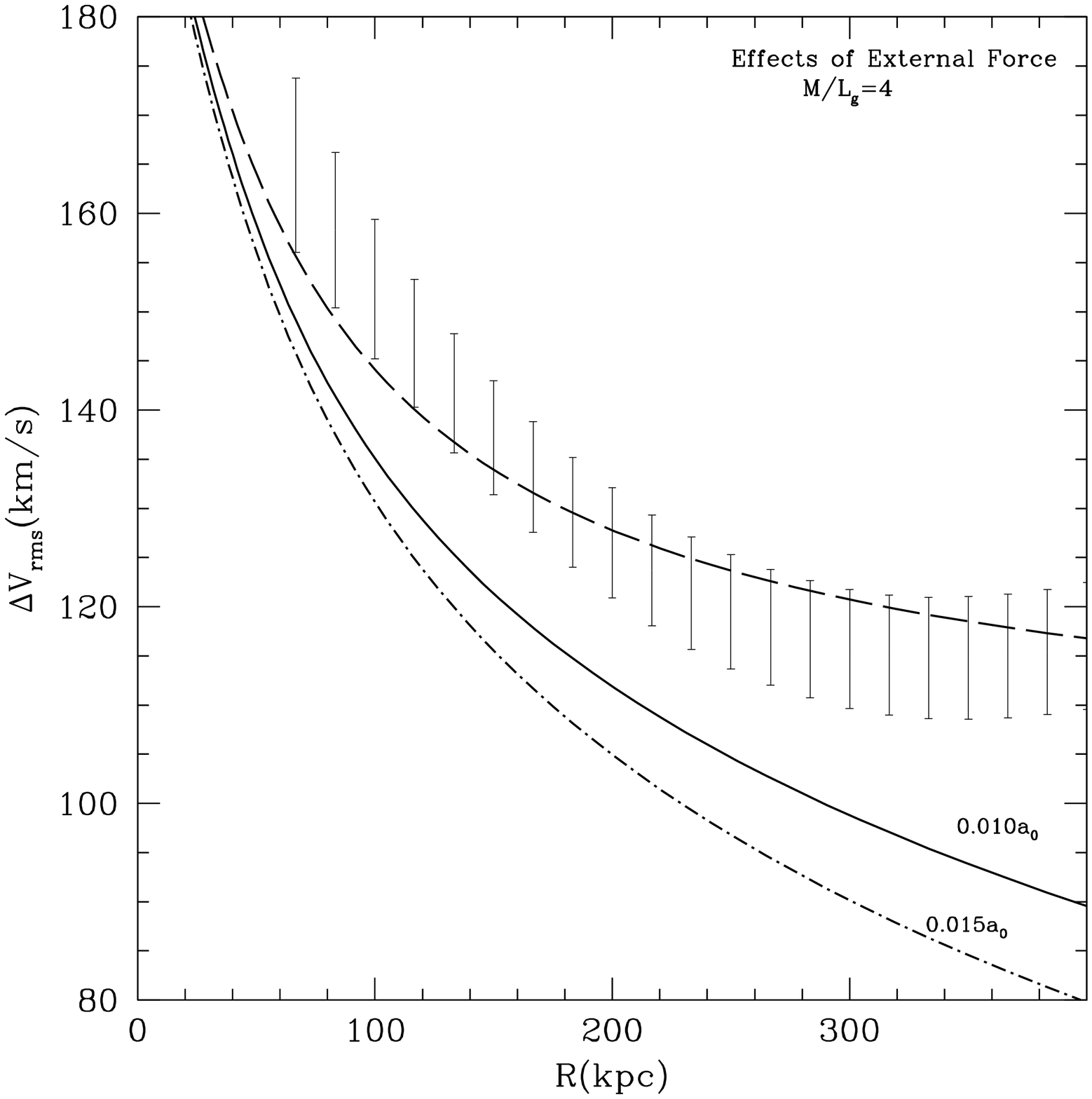}{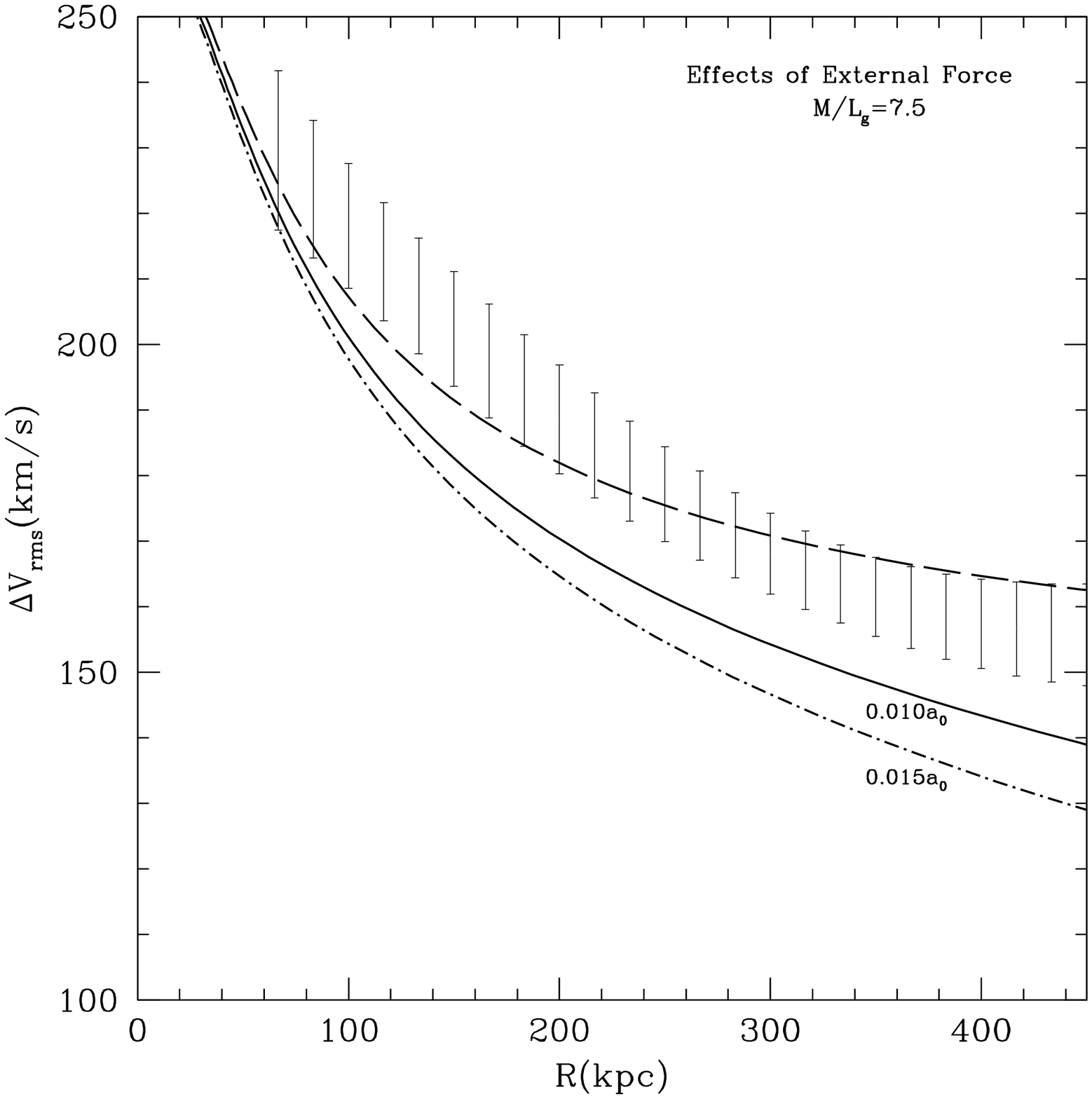}
\caption{Effects of external gravity force. Parameters are the same as in
Figure~3. For both low (left panel) and high (right panel) luminosity
bins we assume the most favorable MOND parameters: radial orbits and
adjusted M/L (models shown with dashed curves in
Figure~\ref{fig:Tuned}). The magnitude of the external field is shown
in the left panel. Models cannot fit the data without additional
increase of M/L.
\label{fig:TunedExt}
}
\end{figure*}

The external gravity force $g_{\rm ext}$ is another MOND component. It
must exist on the level of $g_{\rm ext}\approx
0.01a_0$. Figure~\ref{fig:TunedExt} shows what happens to the models
when we add the external force. As the starting models we use the best
fits shown in the Figure~\ref{fig:Tuned} as dashed curves. The models
with the external force make the fits much worse. Taken at face value,
the models with realistic $g_{\rm ext}=0.015a_0$ can be rejected.
Again, we can make the model work if we increase M/L by a factor $\sim
1.5$. Yet, this will make the situation with stellar masses even
worse, than what we already have.

\section{Discussion and Conclusions}
\label{sec:conclusions}
Using the SDSS DR4 data we study the distribution and velocities of
satellites orbiting red isolated galaxies. We find that the surface
number-density of the satellites declines almost as a power law with
the slope $-2.5-3$. The distribution of the line-of-sight velocities
is nearly a perfect  Gaussian distribution with a constant component
due to interlopers. The rms velocities are found to gradually decline
with the projected distance. The constant rms velocity (isothermal
solution) can be rejected at a 10~$\sigma$ level.

Observational data strongly favor the standard cosmological model: all
three major statistics of satellites -- the number-density profile,
the line-of-sight velocity dispersion, and the distribution function
of the velocities -- agree remarkably well with the theoretical
predictions. Thus, the success of the standard model extends to scales
(50-500)~kpc, much lower than what was previously considered.

MOND fails badly in cases with any acceptable power- law approximation
for the number-density of satellites and a constant velocity
anisotropy by producing sharply declining velocities at small
distances followed by nearly flat velocities at large distances --
just the opposite of what is observed in real galaxies. Models may be
made to fit the satellites data only when all the following conditions
are fulfilled:

\begin{enumerate}
\item The slope of the density in the central 50~kpc is less then -2.
\item Satellite velocities are nearly radial in outer regions ($\beta\approx
  0.6-0.7$ at $R=200-300$~kpc).
\item Mass-to-light ratio increases from $M/L_g\approx 3.5$ at
  $M_g=-20.9$ to $M/L_g\approx 12$ ($M/L_B\approx 17$) at $M_g=-22$.
\item External force of gravity is smaller than the expected value of $g_{\rm ext}$ $=0.015a_0$.
        The negative effect of the external force can be removed by 
        additionally increasing the M/L ratios. 
\end{enumerate}

We find that the later three conditions are difficult to realize in
nature.  Satellites do not fall in to their parent galaxies with zero
tangential velocities.  The velocities are induced by other
neighboring galaxies and by large-scale structures such as
filaments. To some degree, it is similar to the tidal torque, which is
responsible for the origin of the angular momentum of galaxies. Yet,
the interactions are more efficient for providing random velocities.
Measurements of peculiar velocities of galaxies the Local Volume
(3-5~Mpc around the Milky Way) found deviations from the Hubble flow
that about 70-80~km/s \citep{Maccio2005}. At these distances the gravitational pull of
other large galaxies in the aria is larger than that of the Milky Way
and we expect that the same deviations exist for the perpendicular
velocity components. When a satellite with 70-80~km/s falls falls from
1~Mpc and gets to 400~kpc from MW its tangential velocity increases
few times (conservation of the angular momentum). So, we expect $\beta
<0.5$. We have the same measurements for the satellites in SDSS. When
we look at distances 0.8--1~Mpc, we find that the rms velocities
hardly correlate with the luminosity and are about 100~km/s. Note that
\LCDM perfectly fits the constraint: at these distances it predicts
$\beta<0.4$. Thus, MOND assumption of very radial orbits does not
agree with the observations.

The large M/L ratios is another problem for MOND. So far M/L was
treated almost as a free parameter \citep{Sanders07,Angus2007}.  This
is not correct. It should be reminded that observationally there are
no indications of large variability of IMF for vastly different stellar
ages and metallicities. The IMF shows similar flattening for masses
smaller than $M_{\odot}$ in the solar neighborhood and in the spheroid
\citep{Chabrier2001}, in the galactic bulge \citep{Zoccali03}, and in
the Ursa Minor dwarf spheroidal galaxy \citep{Feltzing99}.  The
galactic bulge has parameters similar to the red galaxies studied in
this paper: very old (on average) stellar population with nearly solar
metallicity.  \citet{Zoccali03} give the $M/L_V=3.67$, which
corresponds to $M/L_g=3$. This is close to what we found for the red
galaxies in SDSS. 

In the case studied here, MOND requires that M/L changes by a factor 2-2.5 when
there is no change in colors and the absolute magnitude changes only
by $\Delta M\approx 0.5$. There is no justification for this to
happen. The situation is similar to what \citet{Sanders07} found for
early type large spirals where MOND models fail for any IMF discussed
in recent 20 years. Only the Salpeter IMF gives better results. Even
in this case the models deviate by a factor of two around the stellar
population predictions. The combination of all these problems makes
MOND a very implausible solution for the observational data on
satellites of galaxies.

There is one significant difference between MOND and \LCDM.  The latter
makes predictions for the distribution of mass and velocities for
isolated galaxies and those predictions match the observational
data.  There are no theoretical predictions for
MOND. What we conveniently called ``MOND predictions'' were actually
{\it requirements}. For example, MOND must produce the radial
velocities of satellites or shallow slope of the number-density in the
central region.   

\section*{Acknowledgments}
We thank S.~McGaugh for challenging us to do MONDian analysis of the
satellite motion.  We are grateful to J.~Holtzman for numerous
comments and suggestions and thank J. Betancort-Rijo for comments.  We
are especially grateful to Angus et al for providing us with the draft
of their paper and for extensive discussions. We still have different
ways of looking at the situation with MOND. Still, these productive
discussions resulted in improvements in their and our papers.  We
acknowledge support by the NSF grant AST-0407072 to NMSU and thank the
Spanish MEC under grant PNAYA 2005-07789 for their support.  Computer
simulations used in this research were conducted on the Columbia
supercomputer at the NASA Advanced Supercomputing Division and on
Seaborg at the National Energy Research Scientific Computing Center
(NERSC).

\end{document}